\documentclass[conference,a4paper]{IEEEtran}
\usepackage[pdftex]{graphicx}
\usepackage{amsmath,amssymb}

\def\(({\left(}
\def\)){\right)}
\def\[[{\left[}
\def\]]{\right]}

\newcommand{\be}{\begin{equation}}
\newcommand{\ee}{\end{equation}}
\newcommand{\bea}{\begin{eqnarray}}
\newcommand{\eea}{\end{eqnarray}}

\renewcommand{\d}{\text {d}}

\let\a=\alpha     \let\d=\delta

 \let\D=\Delta   
         
 \let\ee=\epsilon \let\r=\rho

\def\MM{{\cal M}}

\def\DD{{\cal D}}\def\AA{{\cal A}}

\def\ol{\overline}

\def\to{\rightarrow}

\newcommand{\beq}{\begin{equation}}
\newcommand{\eeq}{\end{equation}}

\begin{document}

\sloppy

\title{On Convergence of Approximate Message Passing}

\author{
\IEEEauthorblockN{Francesco Caltagirone, Lenka
    Zdeborov\'a}  \IEEEauthorblockA{Institut de Physique Th\'eorique\\
    CEA Saclay and URA 2306, CNRS\\ 91191 Gif-sur-Yvette,
    France.}
\and \IEEEauthorblockN{Florent Krzakala} \IEEEauthorblockA{
    Laboratoire de Physique Statistique, 
\'Ecole Normale Sup\'erieure\\
    and Universit\'e Pierre et Marie Curie,
    Rue Lhomond Paris 75005  France\\
    ESPCI and CNRS UMR 7083, 10 rue Vauquelin,
 Paris 75005 France}
}


\maketitle

\begin{abstract}
  Approximate message passing is an iterative algorithm for compressed sensing and related
  applications. A solid theory about the performance and convergence
  of the algorithm exists for measurement matrices having iid entries
  of zero mean. However, it was observed by several authors that for
  more general matrices the algorithm often encounters convergence
  problems. In this paper we identify the reason of the
  non-convergence for measurement matrices with iid entries and
  non-zero mean in the context of Bayes optimal inference. Finally we
  demonstrate numerically that when the iterative update is changed
  from parallel to sequential the convergence is restored.
\end{abstract}

\section{Introduction}

Approximate message passing
\cite{DonohoMaleki09,DonohoMaleki10,Rangan10b} is an algorithm derived
from belief propagation that has been recently used with success in a
number of sparse estimation problems, see
e.g. \cite{schniter2010turbo,som2012compressive}. Highly non-trivial
theoretical results were obtained on the performances of this
algorithm
\cite{BayatiMontanari10,KrzakalaPRX2012,DonohoJavanmard11}. Based on
these developments and the promising nature of their results we can
anticipate that AMP based algorithms will become the state-of-the-art
algorithms for many problems of practical interest.

Just as with any iterative algorithm the main question about AMP,
besides its performance, is its convergence. This question is largely
open except for the case of compressed sensing, i.e. estimation of a
sparse ${\bf x}$ from noisy linear projections
\beq {\bf y} = F {\bf x} + {\bf \xi} \, \eeq 
with matrices $F$ having iid entries of zero mean, and ${\xi}$ a white
Gaussian noise of variance $\Delta$. This last case has been treated
in the rigorous proofs in the very large signal size limit of
\cite{BayatiMontanari10,bayati2012universality}. However, for many
other sparse estimation problems, or for slightly more general
matrices $F$, the basic version of AMP fails to converge (and worst,
can diverge violently). Attempts to fix these convergence issues were
so far limited to rather basic and empirical strategies such as
damping the iterations in various ways, or transforming the matrix by
subtracting its mean.
Such strategies are
rarely discussed in the literature and often appear only in the
associated implementations available online.
Moreover, they are far from ensuring the convergence in all cases and
some of these strategies (e.g. the mean removal) are not usable in more
challenging signal processing settings where approximate message
passing can be applied (e.g. the dictionary learning problem
\cite{krzakala2013phase}). The main motivation of this work is to
understand the origin of some of these convergence problems.

The structurally simplest case where AMP fails to converge appears to
be when the measurement matrix $F$ has iid entries of non-zero
mean. This problem was noticed by several authors,
e.g. \cite{Rangan10b,KrzakalaMezard12}, and fixed in the
implementations by removing the mean of the matrix. Indeed, the
average of element of the measurement vector ${\bf y}$ reads
 \beq
\overline { y} = \frac{1}{M}\sum_\mu y_\mu = \sum_i \left( \frac{1}{M}\sum_\mu F_{\mu i}\right)
x_i \, .
\eeq 
We denote $\overline F_i= \sum_\mu F_{\mu i}/M$ the average value of
$F$ for column $i$.  One can then work with the modified system
$y_{\mu} - \overline y= \sum_i (F_{\mu i}-\overline F_i) x_i$ where
the mean of the new sensing matrix $F_{\mu i}-\overline F_i$ is zero. A similar (but
different) trick is used in the implementation of
\cite{Rangan10b}. This "remove mean'' strategy is, however, not fully
satisfactory because it is not understood why it is needed in the
first place, nor under what conditions it restores the
convergence. Moreover in some more general settings it is not
applicable at all.


The goal of this paper is to analyze the origin of the
non-convergence for non-zero mean matrices and discuss general
strategies to prevent it. Such an understanding is a step towards the
design of robustly convergent and hence more efficient AMP-based
algorithms. We will hence consider matrices with entries generated as
follows 
\beq F_{\mu i } = \frac{\gamma}{N} + \frac{1}{\sqrt{N}} {\cal
  N}(0,1) \, . \ \eeq 
For $\gamma = 0$ this is the case that has been considered in the
literature. To be specific and simple we will consider that the signal
${\bf x}$ was generated to have $\rho N$ non-zero entries that are iid
normally distributed with zero mean and unit variance \beq P(x) =
(1-\r)\d(x)+\r {\cal N}(0,1) \, . \label{eq:Px} \eeq 
We will consider the Bayesian version of the AMP algorithm that uses
this prior information about the signal. A first observation is that
AMP does not depend on $\gamma$ in an explicit way: this can be
checked explicitly by repeating the detailed derivations of AMP
present in the literature for $\gamma>0$ (follow e.g. the derivation
in \cite{KrzakalaMezard12}).

On the other hand the asymptotic analysis of the performance of the
algorithm --- the state evolution
\cite{DonohoMaleki09,BayatiMontanari10}--- depends on $\gamma$
explicitly and hence we have to rederive it. The analysis of the
state evolution for $\gamma>0$ will lead to an
understanding of the origin of the convergence problems.

\section{The AMP algorithm}
\label{algo}
We consider the AMP algorithm in the form that was derived in
\cite{DonohoMaleki10,Rangan10b,KrzakalaMezard12}. The main steps are
a) going from belief propagation (BP) to a relaxed BP (r-BP) where
only the two first moments of all messages are kept and b) using $N$
sites marginals instead of $N \times M$ messages and adding the
compensating Onsager terms 
\cite{thouless1977solution}. Finally, AMP reads:
\begin{align}
V^{t+1}_\mu &= \sum_i F_{\mu i}^{2} \, v^{t}_i \, , \label{TAP_ga} \\
\omega^{t+1}_\mu &= \sum_i F_{\mu i} \, a^t_i -\frac{
  (y_\mu-\omega^t_\mu)}{\Delta +V^t_\mu} \sum_i F_{\mu i}^2\,
v^t_i \, , \label{TAP_al}  \\
(\Sigma^{t+1}_i)^2&=\left[ \sum_\mu \frac{F^2_{\mu i}}{\Delta +
    V^{t+1}_\mu} \right]^{-1}\, ,
      \label{TAP_U}\\
      R^{t+1}_i&= a^t_i + \frac{\sum_\mu F_{\mu i} \frac{(y_\mu - \omega^{t+1}_\mu)}{\Delta
        + V^{t+1}_\mu}}{ \sum_\mu \frac{ F_{\mu i}^2}{\Delta + V^{t+1}_\mu}}\, , \label{TAP_V}\\
      a^{t+1}_i &=   f_1\left((\Sigma^{t+1}_i)^2,R^{t+1}_i\right) ,  \label{TAP_a}\\
      v^{t+1}_i &= f_2\left((\Sigma^{t+1}_i)^2,R^{t+1}_i\right) \, . \label{TAP_v}
\end{align}
where $f_k(\Sigma^2,R)$, here and in what follows, are the $k$-th
connected cumulants w.r.t. the probability measure 
\beq
{\cal Q}(x)=\frac{1}{Z(\Sigma^2,R)}P(x)\frac{e^{-\frac{(x-R)^2}{2\Sigma^2}}}{\sqrt{2\pi
    \Sigma^2}} \, , \label{eq:nu}
\eeq
where $Z(\Sigma^2,R)$ is the normalization constant. 

The variables $a_i$ and $v_i$ are the AMP estimators for the mean and
variance of the component $i$ of the signal. The quality of the
reconstruction can be evaluated by computing the mean squared error
(MSE) 
\beq
E^t=\frac{1}{N}\sum_{i=1}^N (s_i - a^t_i)^2 \eeq and the average
variance \beq V^t = \frac{1}{N}\sum_{i=1}^N v_i \, .  \eeq

When $\gamma\!=\!0$, the performance of the AMP algorithm was analyzed
rigorously in the limit of large system size via the state evolution
$(E^{t+1}, V^{t+1}) = G(E^t,V^t)$, where $G$ is a function specified
in
\cite{BayatiMontanari10,DonohoMaleki10,Rangan10b,KrzakalaMezard12}. An
important property of the Bayes optimal inference (i.e. when the
signal was indeed generated from the assumed prior distribution) is
that the two paramaters are equal in the large size limit, $E^t\!=\!V^t$, and the state evolution hence
reduces to an iterative equation of a single real number, which is
amenable to rigorous analysis \cite{BayatiMontanari10}. In statistical
physics $E^t\!=\!V^t$ is called the Nishimori condition and is
discussed in the context of compressed sensing in detail in
\cite{KrzakalaMezard12}. In general, when $\gamma\!=\!0$ we observed by
analyzing the state evolution equations that even when at initial times
$E^{t=0}\!\neq\! V^{t=0}$ the equality $E^t\!=\!V^t$ is restored after
a sufficient number of iterations.

\section{State evolution with non-zero mean matrices} 

The state evolution of the AMP algorithm can be derived for
measurement matrices with non-zero mean $\gamma>0$.  Here we follow closely the derivation and
notation from \cite{KrzakalaMezard12} for zero mean matrices. 
Among the different variables, the statistical distribution of $R_i$ plays a crucial role in the determination of the state
evolution it can be written as 
\beq
R^t_i=s_i+\frac{1}{\a}r^t_i\, ,
\eeq
where $s_i$ is the original signal component and
\beq
r_i^t=\sum_\mu F_{\mu i} \xi_\mu + \sum_\mu F_{\mu i} \sum_{j\neq i} F_{\mu j} (s_j -
a_{j\to \mu}^t)
\eeq
is a Gaussian random variable, and $a_{j\to \mu}^t$ is an auxiliary
variable related closely to $a_i^t$ that appears in the derivation of
the AMP algorithm. Assumptions used to derive AMP can be used to
compute the  mean and variance of  $r_i^t$ over realizations of the
problem. In the leading order we get   
\bea
\ol{r^{t}}&=&\a \gamma^2 D^t \, , \\
\text{var}(r^t)&=& \a (E+\D + \gamma^2 D^2) \, ,
\eea
where we have defined a new order parameter 
\beq
D^t=\frac{1}{N} \sum_j (s_j - a^t_j) \, .
\eeq
The parameter $D^t$ is not needed for zero mean matrices
$\gamma=0$. For $\gamma>0$, however, the state evolution is written
in terms of three parameters $E^t$, $V^t$ and $D^t$. The remaining
steps in the derivation are basically identical to those for zero
mean matrices and following \cite{KrzakalaMezard12} we obtain 
\begin{align}
V^{t+1}&=\int \, {\rm d}s \, P(s) \, \int \DD z \times \label{eq:V} \\ 
& f_2 \left( \frac{\D + V^t}{\a} , s+z \AA(E^t,D^t)+\gamma^2 D^t \right)\,
 ,   \nonumber \\
E^{t+1}&=\int \, {\rm d}s \, P(s) \, \int \DD z \times \label{eq:E} \\
 &\left[s-  f_1 \left( \frac{\D + V^t}{\a} , s+z \AA(E^t,D^t)+\gamma^2 D^t
   \right)  \right]^2\, , \nonumber \\
D^{t+1}&=\int \, {\rm d}s \, P(s) \, \int \DD z \times \label{eq:D} \\
 &\left[ s - f_1 \left( \frac{\D + V^t}{\a} , s+z \AA(E^t,D^t)+\gamma^2
     D^t \right)  \right]\, .  \nonumber 
\end{align}
where $\DD z$ is a Gaussian measure and 
\beq
\AA(E^t,D^t)=\sqrt{\frac{E^t + \D + \gamma^2(D^t)^2 }{\a}}\, .
\eeq
When the mean of the measurement matrix is zero,  $\gamma=0$, these
equations clearly reduce to those derived in \cite{Rangan10b,KrzakalaMezard12}. 

Also for $\gamma>0$ we can identify the Nishimori condition, which
reads $E^t = V^t$ (for the same reasons as for the previous case) and
$D^t = 0$ (since under Bayes optimal inference the mean of the
estimator must be equal to the true mean of the signal). It is a question of simple algebraic
verification to see that starting with $E^t = V^t$ and $D^t = 0$
eqs. (\ref{eq:V}-\ref{eq:D}) lead to $E^{t+1} = V^{t+1}$ and $D^{t+1}
= 0$. Hence if we restrict ourselves to the space on which the
Nishimori conditions hold (called the Nishimori line) there is no
difference between the $\gamma=0$ and $\gamma>0$ case.

\section{Instability of the Nishimori line} 
\label{stab}
In this Section we analyze the dynamical stability of the Nishimori
line (NL) under iterations of eqs. (\ref{eq:V}-\ref{eq:D}).
We consider the space $(K,D)$ orthogonal to the NL, where $K=V-E$.
We know that in this space $(K^*=0,D^*=0)$ is a fixed point. We can generically write 
\beq
\begin{split}
K^{t+1}&=f_K(V^t,K^t,D^t)\, ,\\
D^{t+1}&=f_D(V^t,K^t,D^t)\, .
\end{split}
\eeq 
To analyze the stability we linearize around the fixed point considering the perturbations
$\d K^t=K^t-K^*$ and $\d D^t=D^t-D^*$. The linearized formula reads
\beq
\left(
\begin{array}{c}
\d K^{t+1}\\
\d D^{t+1}\\
\end{array}
\right) 
=\MM \cdot
\left(
\begin{array}{c}
\d K^{t}\\
\d D^{t}\\
\end{array}
\right) 
\eeq
with
\beq
\MM= \left(
\begin{matrix}
\partial _K f_K(V^t,K^*,D^*)& \partial _D f_K(V^t,K^*,D^*)\\
\partial _K f_D(V^t,K^*,D^*)&  \partial _D f_D(V^t,K^*,D^*)
\end{matrix}
\right) \, . \label{eq:M}
\eeq

It follows from a straightforward algebraic verification that both the
off-diagonal terms (the cross derivatives) are zero for the
distribution $P(x)$ from eq.~(\ref{eq:Px}). The matrix ${\cal M}$
(\ref{eq:M}) is hence diagonal.  For a more generic prior
distribution the situation is slightly more involved, but
qualitatively analogous to the one of (\ref{eq:Px}). The diagonal
terms read 
\begin{align}
\partial_D f_D(V^t)&=-\frac{\a \gamma^2}{\D + V^t} \int \, ds \, P(s) \int
\DD z \, f_2\left( A^2,s+zA \right) \nonumber \\
&=-\frac{\a \gamma^2 V^{t+1}}{\D + V^t} \, ,\label{eq:DD} \\
\partial_K f_K(V^t)&=-\frac{1}{2}\frac{1}{\D + V^t} \int \, ds \, P(s) \int \DD z \, \left\lbrace f_4\left( A^2,s+zA \right) \right. \nonumber \\ 
&\left.+2(f_2\left( A^2,s+zA \right) )^2 \right. \label{eq:KK} \\
&\left.+ 2 \left[ f_1\left( A^2,s+zA \right)-s \right] f_3\left(
    A^2,s+zA \right) \right\rbrace \, , \nonumber
\end{align}
where, as before, the functions $f_k(\Sigma^2,R)$ are the $k$-th
connected cumulants with respect to the measure ${\cal Q}(\Sigma^2,R)$
(\ref{eq:nu}), and where we denoted 
\beq
A \equiv \sqrt{\frac{\D + V^t}{\a}} \, .\label{eq:A}
\eeq
The term
$\partial_K f_K(V^t)$ is independent of $\gamma$ and its module is
always smaller than one. Hence the Nishimori line is stable in the
direction $K=V-E$. 

\begin{figure}
\centering
\includegraphics[scale = 0.6]{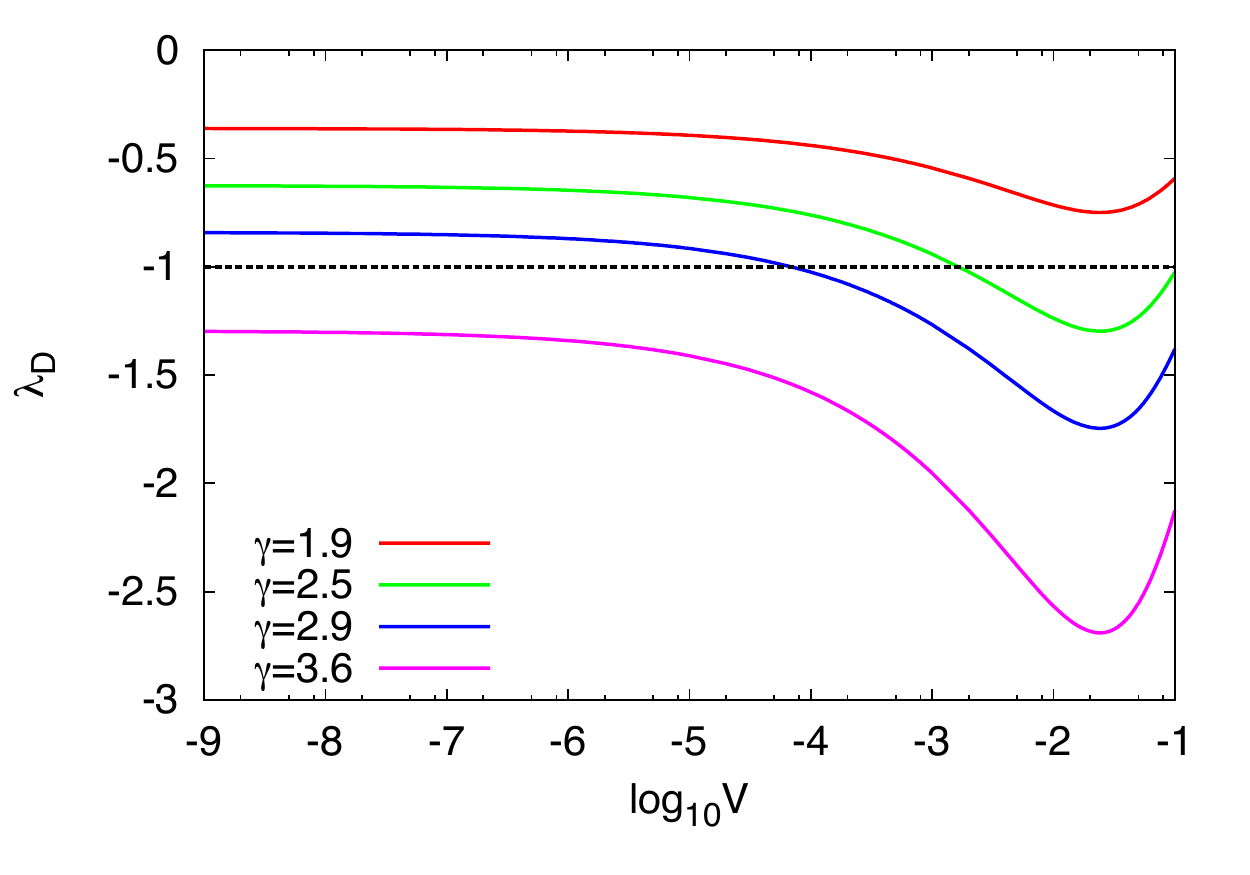}
\caption{The term $\lambda_D = \partial_D f_D(V^t)$ associated to the
  stability of the Nishimori line in the $D$-direction as a function of
  the MSE for $\r=0.1$, $\D=10^{-10}$ and $\a=0.3$. Three different
  regimes can be identified, one in which $|\lambda_D|$ is always less than
  $1$, the second in which $|\lambda_D|$ is larger than $1$ in a region, and
  the third in which $|\lambda_D|$ is larger than one along the whole
  Nishimori line down to the fixed point. The critical values (as defined in the text) for this case are
  $\gamma_c^{(1)}\simeq 2.197$, $\gamma_c^{(2)}\simeq 3.162$. }
\label{plot_eigen}
\end{figure}

\begin{figure}
\centering
\includegraphics[scale = 0.6]{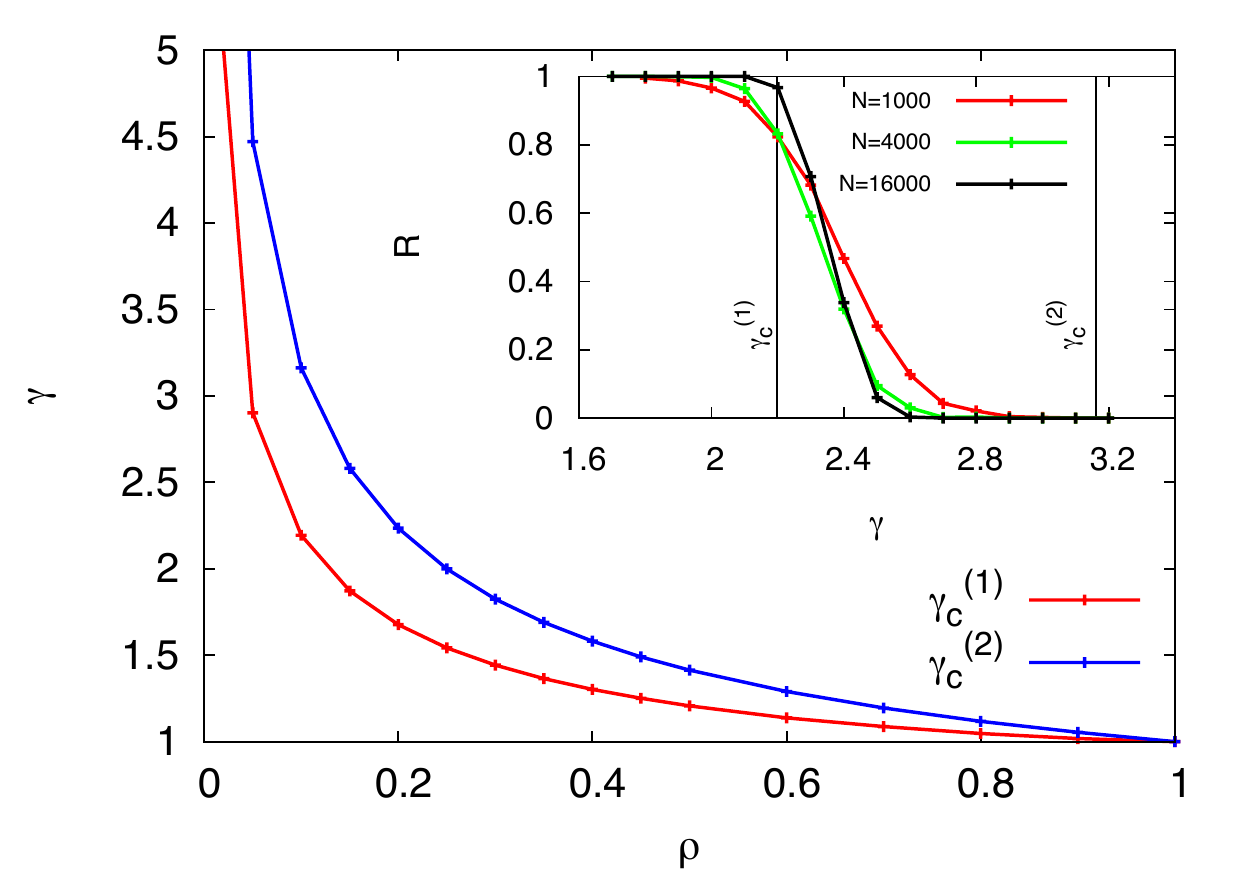}
\caption{[Main frame] The threshold values for the mean of the measurement matrix
  above which the state evolution on the Nishimori line (i.e. $E=V$
  and $D=0$) is not stable. Above $\gamma_c^{(1)}$ only part of the line
  is unstable, above $\gamma_c^{(2)}$ the full line is unstable. For zero
  measurement noise these values do not depend on the undersampling
  rate $\alpha$. For weak measurement noise only the line
  $\gamma_c^{(2)}$ depends weakly on both $\Delta$ and $\alpha$. 
[Inset] The convergence rate $R$ of the AMP algorithm as a function
  of the mean of the measurement matrix $\gamma$ with $\r=0.1$, $\D=10^{-10}$ and $\a=0.3$ for 
different values of the signal size $N$. We can see that the transition is close to the first critical value $\gamma_c^{(1)}$ (marked by the 
vertical line on the left) and it is smoother for low $N$ and sharper for larger $N$. For very large 
$N$ we also expect the transition to move towards the vertical line on
the right ($\gamma=\gamma_c^{(2)}$), but this effect is not visible at the $N$ we are able to reach.}. 
\label{lam_crit}
\end{figure}

On the other hand the term $\lambda_D = \partial_D f_D(V^t)$ has a non-trivial
behavior that we illustrate in Fig.~\ref{plot_eigen} for $\r=0.1$,
$\a=0.3$, $\D=10^{-10}$ and, respectively, $\gamma=1.9$, $\gamma=2.5$, $\gamma=2.9$ and $\gamma=3.6$. In the figure we identify three different regimes: 
\begin{itemize}
\item For $|\gamma|<\gamma_c^{(1)}$ the eigenvalue $\lambda_D$ is always less than $1$ in modulus. 
\item For $\gamma_c^{(1)}<|\gamma|<\gamma_c^{(2)}$ the eigenvalue becomes greater
  than $1$ in modulus in a certain portion of the Nishimori line. 
In this region the evolution tends to make $|D|$ larger, while at the same time $V$ and $E$ decrease.  
\item For $|\gamma|>\gamma_c^{(2)}$ the eigenvalue $\lambda_D$ is larger than $1$
  in modulus in the whole range down to the fixed point. This tells us 
that any fluctuation of $D$ will be progressively enhanced.
\end{itemize}
We further realize that the expression used to calculate
$\lambda_D$ depends on the value $V$ only through the variable $A$
(\ref{eq:A}) and not in an explicit way on the parameters $\alpha$ and
$\Delta$. This means that the threshold value $\gamma_c^{(1)}$ is from its
definition independent of $\alpha$ and $\Delta$. The threshold value
$\gamma_c^{(2)}$ is also independent of $\alpha$ for $\Delta=0$ and only
weakly dependent on both $\alpha$ and $\Delta$ for small values of
$\Delta$. In Fig.~\ref{lam_crit} we hence plot the two threshold
values for $\Delta=0$ (in which case they are both independent of
the undersampling $\alpha$) as a function of the sparsity $\rho$. 

\section{Comparing state evolution to AMP}
\label{behavior}

We now discuss how does the instability of the Nishimori
line translate into the behavior of the state evolution (SE) initialized usually as $E^{t=0}=V^{t=0}=\rho$ (corresponding to $a^{t=0}_i=0$ and
$v^{t=0}_i=\rho$) and $D^{t=0}=0$. The SE was derived to correspond to the
behavior of the AMP algorithm for sufficiently large system sizes
$N$. We observe that
\begin{itemize}
\item For $|\gamma|<\gamma_c^{(1)}$ the SE converges to the fixed point with monotonically decreasing $E=V$. There are 
really infinitesimal fluctuations in $D$ that are due to numerical precision but they are harmless.  
\item For $\gamma_c^{(1)}<|\gamma|<\gamma_c^{(2)}$ the SE converges to the fixed point with monotonically decreasing $E=V$. 
In the region of $V$ in which $|\lambda_D|$ is larger than one, we observe that the numerical fluctuations of $D$ are slightly 
increased (especially if we are close to $\gamma_c^{(2)}$), without
changing qualitatively the behavior of $V$, and when $|\lambda_D|$ becomes again smaller than 1 the fluctuations are reabsorbed. 
\item For $|\gamma|>\gamma_c^{(2)}$ the fluctuations of $D$ are
  increased along the whole line $E=V$. At some point these
  fluctuations reach so large values that the difference $K=E-V$ grows
  and we observe a divergence of both $E$ and $V$.
\end{itemize}

Therefore, while with infinite numerical precision the SE should stay on the Nishimori line and 
converge whatever the value of $\gamma$ is, from the practical point of view the fluctuations due to numerical precision 
are sufficient to cause divergence in the third regime. Of course in
the AMP algorithm the typical fluctuations are of order $1/\sqrt{N}$ hence
relatively large and that is the reason why for $|\gamma|>\gamma_c^{(2)}$ AMP
never converges. In fact these finite size fluctuations are so strong
that even in the second regime $\gamma_c^{(1)}<|\gamma|<\gamma_c^{(2)}$ AMP might have
problems. Therefore we observe a smooth transition in the success rate $R=$($\#$successes/$\#$failures) between $\gamma_c^{(1)}$ and $\gamma_c^{(2)}$ for finite $N$. 
When $N$ is increased this smooth transition becomes sharper. In the inset of Fig. \ref{lam_crit} we show the success rate of the AMP algorithm averaged over $1000$ random instances of 
the measurement matrix for $N=1000,4000$ and over $500$ instances for $N=16000$. 
We see that, even if asymptotically, the reference value for the
success/failure transition would be $\gamma_c^{(2)}$, for all practical
system sizes, the right threshold to look at is rather~$\gamma_c^{(1)}$.

\section{Reducing the instability}

There are at least two strategies that appear in the implementations
of the AMP algorithm that improve its convergence. Let us discuss them
now in the context of the above analysis. 

\paragraph{Damping} 

A popular and generic strategy to improve convergence of iterative
algorithms is ``damping'', i.e. in every new iteration we update the
variables only partially. Such damping (with different schemes) appears in basically every
available implementation of AMP. In the view of the preceding
analysis a dynamical instability is mitigated by such damping and the
eigenvalue $|\lambda_D|$ becomes effectively smaller. Indeed AMP with
damping converges well even for matrices with means slightly larger
than those corresponding to $\gamma_c^{(2)}$ in Fig.~\ref{lam_crit}. 

\paragraph{Expectation maximization learning} 

In this paper so far we assumed the prior knowledge of the probability
distribution of signal elements as well as of the measurement noise $\Delta$
and the sparsity $\rho$. A classical strategy of expectation
maximization was suggested, tested and implemented in \cite{KrzakalaPRX2012,VilaSchniter11} in order
to learn these parameters when they are not known apriori. A careful
investigation of the AMP algorithm with EM learning leads to a
conclusion that with the learning the AMP has better convergence
properties than without. 

This can come as a surprise at first, but in the view of our above
investigation it can now be easily explained.  The EM update in a
sense imposes (in an iterative way) the Nishimori condition, see the
derivation of EM in \cite{KrzakalaPRX2012}, hence it should be
expected that it also stabilizes the Nishimori line and consequently
improves the convergence of AMP.

\section{The sequential redemption} 
\begin{figure}
\centering
\includegraphics[scale = 0.6]{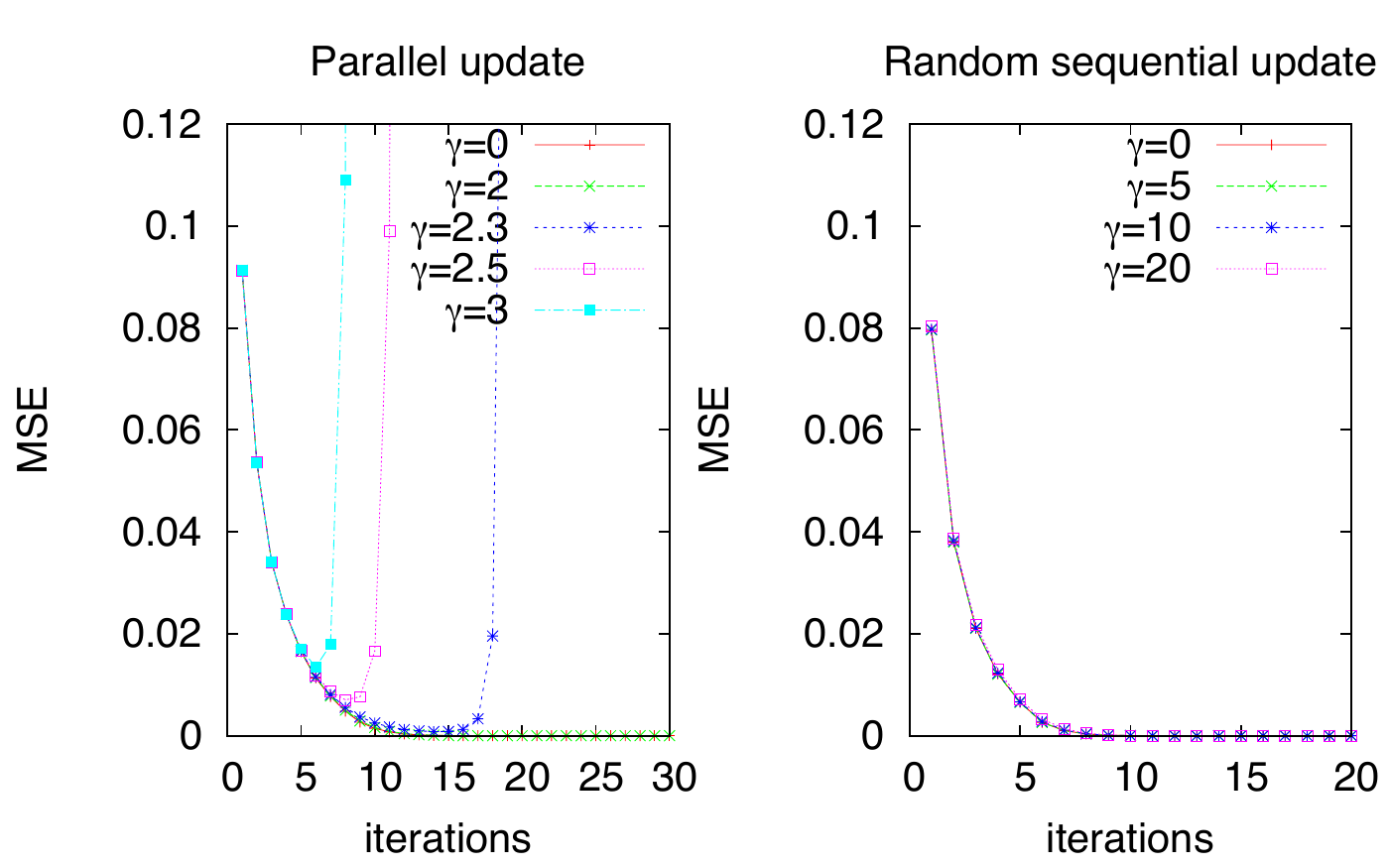}
\caption{A comparison of the effect of a non-zero mean $\gamma$ on
  the relaxed belief propagation algorithm (\ref{A_mu}-\ref{eqq4})
  with different update scheme using
  $\alpha=0.3$, $\rho=0.1$, $\Delta=10^{-10}$ and $N=10^4$. Left:
  Parallel update corresponding to (and equivalent to) AMP. 
For $\gamma>2.3$, even the damped AMP diverges. 
  Right:
  Random sequential update. In that case, r-BP
  converges very fast even for large values of the mean $\gamma$.}
\label{BP_seqvspara}
\end{figure}

AMP being so sensitive to the mean of the matrix elements is
surprising because the standard BP, when applied to discrete random
problems, does not experience such problems.  In this last section we
argue that the convergence problems in the case of CS with non-zero
mean measurement matrices are actually specific to the ``parallel
updates'' (involving only matrix multiplications) performed naturally
in the AMP algorithm that we presented in Sec. \ref{algo}. Let us
recall the so-called relaxed-BP (r-BP) algorithm \cite{Rangan10} (for
present notations see \cite{KrzakalaMezard12}) where messages are sent
on the factor graph:
\bea
A_{\mu\to i} &=& \frac{F^2_{\mu i}}{\Delta + \sum_{j\neq i} F^2_{\mu j} v_{j\to \mu}}  \, , \label{A_mu} \\
B_{\mu \to i} &=& \frac{F_{\mu i}(y_\mu - \sum_{j\neq i} F_{\mu
    j}a_{j\to \mu})}{\Delta + \sum_{j\neq i} F^2_{\mu j} v_{j\to
    \mu}} \, , \label{B_mu} \\
a_{i\to \mu}&=&f_1\left(\frac{1}{\sum_{\gamma\neq \mu}A_{\gamma \to
      i}},\frac{\sum_{\gamma\neq \mu}B_{\gamma \to i}}{
    \sum_{\gamma\neq \mu}A_{\gamma \to
      i}   }\right)\, , \label{eqq1}\\
v_{i\to \mu}&=&f_2\left(\frac{1}{\sum_{\gamma\neq \mu}A_{\gamma \to
      i}},\frac{\sum_{\gamma\neq \mu}B_{\gamma \to i}}{
    \sum_{\gamma\neq \mu}A_{\gamma \to
      i}   }\right)  \, ,\label{eqq2}\\
a_{i}&=&f_1\left(\frac{1}{\sum_{\gamma}A_{\gamma \to
      i}},\frac{\sum_{\gamma}B_{\gamma \to i}}{ \sum_{\gamma}A_{\gamma
      \to
      i}   }\right) \, , \label{eqq3}\\
v_{i}&=&f_2\left(\frac{1}{\sum_{\gamma}A_{\gamma \to
      i}},\frac{\sum_{\gamma}B_{\gamma \to i}}{ \sum_{\gamma}A_{\gamma
      \to i} }\right)\, .  \label{eqq4}\eea 
We intentionally wrote this
algorithm without the time indices, because the update can be
performed in two ways. First in the parallel one where {\it all}
variables are updated at time $t$ given the state at time $t-1$. The
second is the random sequential update where one picks {\it a single}
index $i$ and updates all messages corresponding to it. For r-BP, this
leads to the same computational complexity, however, it is important to
realize that AMP is actually written assuming the r-BP with 
the parallel update. In Fig.~\ref{BP_seqvspara} we compare the
behavior of parallel and random sequential r-BP: as we see, the
sequential update does not seem to be affected by the non-zero mean.

This observation of the parallel update being more problematic than
the sequential one is actually not surprising a posteriori.
In fact, such a lack of convergence is known
to occur in parallel iterations in many problems due to instabilities
just like the one we have studied here (see for instance the 
``modularity" instability in the hard-core model \cite{rivoire2004glass} and
coloring \cite{zdeborova2007phase} problems on random graphs). Using
instead, when possible, a sequential r-BP update is therefore an
interesting alternative.  Nevertheless, it is not a {\it universal}
solution since it by no means guarantees convergence for all
matrices. Also, the disadvantage of the sequential r-BP update is that it
looses the nice property of only involving matrix multiplication, a
crucial property for scalability for operators, such as the fast
Fourier transform, for which there exist efficient multiplication methods.

\section{Conclusions}

We have analyzed the convergence problems of AMP in the specific case
of compressed sensing with measurement matrices having iid entries of
non-zero mean.  Despite the fact that the AMP iterations are not
modified w.r.t. the case of zero mean, the state evolution does
contain an additional order parameter. The main result of the paper,
contained in Sec. \ref{stab}, is that the presence of this third
parameter causes instabilities of the so-called Nishimori line and,
therefore in the algorithm itself, if the mean of the matrix elements
exceeds some critical value. In the last section we show
that the convergence issue for matrices of non-zero mean are strongly
mitigated when random sequential update is used in the message passing
instead of the parallel one that is standard to AMP.

This analysis represents a step towards understanding of the
nature of convergence issues in message passing algorithms that are
ubiquitous in problems ranging from physics to information
theory. More complete understanding of these issues is needed before
message passing algorithms can become part of standard toolbox to
solve a wide range of problems of practical interest.

\section*{Acknowledgment}
This work has been supported by the ERC under the European
Union’s 7th Framework Programme Grant Agreement 307087-SPARCS, and by
the project TASC of
the Labex PALM. 

\bibliographystyle{IEEEtran}
\bibliography{refs}

\begin{thebibliography}{10}
\providecommand{\url}[1]{#1}
\csname url@samestyle\endcsname
\providecommand{\newblock}{\relax}
\providecommand{\bibinfo}[2]{#2}
\providecommand{\BIBentrySTDinterwordspacing}{\spaceskip=0pt\relax}
\providecommand{\BIBentryALTinterwordstretchfactor}{4}
\providecommand{\BIBentryALTinterwordspacing}{\spaceskip=\fontdimen2\font plus
\BIBentryALTinterwordstretchfactor\fontdimen3\font minus
  \fontdimen4\font\relax}
\providecommand{\BIBforeignlanguage}[2]{{%
\expandafter\ifx\csname l@#1\endcsname\relax
\typeout{** WARNING: IEEEtran.bst: No hyphenation pattern has been}%
\typeout{** loaded for the language `#1'. Using the pattern for}%
\typeout{** the default language instead.}%
\else
\language=\csname l@#1\endcsname
\fi
#2}}
\providecommand{\BIBdecl}{\relax}
\BIBdecl

\bibitem{DonohoMaleki09}
D.~L. Donoho, A.~Maleki, and A.~Montanari, ``Message-passing algorithms for
  compressed sensing,'' \emph{Proc. Natl. Acad. Sci.}, vol. 106, no.~45, pp.
  18\,914--18\,919, 2009.

\bibitem{DonohoMaleki10}
D.~Donoho, A.~Maleki, and A.~Montanari, ``Message passing algorithms for
  compressed sensing: I. motivation and construction,'' in \emph{IEEE
  Information Theory Workshop (ITW)}, 2010, pp. 1 --5.

\bibitem{Rangan10b}
S.~Rangan, ``Generalized approximate message passing for estimation with random
  linear mixing,'' in \emph{IEEE International Symposium on Information Theory
  Proceedings (ISIT)}, 2011, pp. 2168 --2172.

\bibitem{schniter2010turbo}
P.~Schniter, ``Turbo reconstruction of structured sparse signals,'' in
  \emph{Information Sciences and Systems (CISS), 2010 44th Annual Conference
  on}.\hskip 1em plus 0.5em minus 0.4em\relax IEEE, 2010, pp. 1--6.

\bibitem{som2012compressive}
S.~Som and P.~Schniter, ``Compressive imaging using approximate message passing
  and a markov-tree prior,'' \emph{Signal Processing, IEEE Transactions on},
  vol.~60, no.~7, pp. 3439--3448, 2012.

\bibitem{BayatiMontanari10}
M.~Bayati and A.~Montanari, ``The dynamics of message passing on dense graphs,
  with applications to compressed sensing,'' \emph{IEEE Transactions on
  Information Theory}, vol.~57, no.~2, pp. 764 --785, 2011.

\bibitem{KrzakalaPRX2012}
F.~Krzakala, M.~M{\'e}zard, F.~Sausset, Y.~Sun, and L.~Zdeborov{\'a},
  ``Statistical physics-based reconstruction in compressed sensing,''
  \emph{Phys. Rev. X}, vol.~2, p. 021005, 2012.

\bibitem{DonohoJavanmard11}
D.~L. Donoho, A.~Javanmard, and A.~Montanari, ``Information-theoretically
  optimal compressed sensing via spatial coupling and approximate message
  passing,'' in \emph{Proc. of the IEEE Int. Symposium on Information Theory
  (ISIT)}, 2012.

\bibitem{bayati2012universality}
M.~Bayati, M.~Lelarge, and A.~Montanari, ``Universality in polytope phase
  transitions and message passing algorithms,'' \emph{arXiv preprint
  arXiv:1207.7321}, 2012.

\bibitem{krzakala2013phase}
F.~Krzakala, M.~M{\'e}zard, and L.~Zdeborov{\'a}, ``Phase diagram and
  approximate message passing for blind calibration and dictionary learning,''
  \emph{arXiv preprint arXiv:1301.5898}, 2013.

\bibitem{KrzakalaMezard12}
F.~Krzakala, M.~M\'ezard, F.~Sausset, Y.~Sun, and L.~Zdeborov\'a,
  ``Probabilistic reconstruction in compressed sensing: Algorithms, phase
  diagrams, and threshold achieving matrices,'' \emph{J. Stat. Mech.}, p.
  P08009, 2012.

\bibitem{thouless1977solution}
D.~Thouless, P.~Anderson, and R.~Palmer, ``Solution of'solvable model of a spin
  glass','' \emph{Philosophical Magazine}, vol.~35, no.~3, pp. 593--601, 1977.

\bibitem{VilaSchniter11}
J.~P. Vila and P.~Schniter, ``Expectation-maximization bernoulli-gaussian
  approximate message passing,'' in \emph{Proc. Asilomar Conf. on Signals,
  Systems, and Computers (Pacific Grove, CA)}, 2011.

\bibitem{Rangan10}
S.~Rangan, ``Estimation with random linear mixing, belief propagation and
  compressed sensing,'' in \emph{Information Sciences and Systems (CISS), 2010
  44th Annual Conference on}.\hskip 1em plus 0.5em minus 0.4em\relax IEEE,
  2010, pp. 1--6.

\bibitem{rivoire2004glass}
O.~Rivoire, G.~Biroli, O.~C. Martin, and M.~M{\'e}zard, ``Glass models on bethe
  lattices,'' \emph{The European Physical Journal B-Condensed Matter and
  Complex Systems}, vol.~37, no.~1, pp. 55--78, 2004.

\bibitem{zdeborova2007phase}
L.~Zdeborov{\'a} and F.~Krzakala, ``Phase transitions in the coloring of random
  graphs,'' \emph{Physical Review E}, vol.~76, no.~3, p. 031131, 2007.

\end{thebibliography}

\end{document}